\documentclass[]{acta}
\usepackage{supertabular,lscape,epsfig}
\usepackage{amssymb}
\usepackage{amsmath}
\usepackage{graphicx}

\SetPages{0}{0}

\SetVol{00}{2023}

\begin{document}

\setcounter{page}{1}

\begin{Titlepage}
\Title{Does the A-type Metallic-line Star IW Persei\\
  Have Non-Uniform Chemical Anomaly on the Surface?}
\Author{Y.~~T~a~k~e~d~a}{11-2 Enomachi, Naka-ku, Hiroshima-shi, Japan 730-0851\\
e-mail: ytakeda@js2.so-net.ne.jp}

\Received{May 29, 2023}
\end{Titlepage}

\Abstract{
IW Per, a single-lined spectroscopic binary with a short period 
of 0.92~d, is known to be a A-type metallic-line (Am) star 
showing anomalous line strengths of specific elements.  
Previously, Kim (1980) reported that its equivalent widths of 
Ca~{\sc ii} 3934, Sr~{\sc ii} 4215, and Sc~{\sc ii} 4320 lines 
(important key lines characterizing the Am anomaly) show cyclic 
variations in accordance with the rotation phase, implying that 
the chemical peculiarities on the surface are not uniform but 
of rather patchy distribution, though no trial of reconfirmation 
seems to have been done so far. 
In order to check the validity of this finding, 10 high-dispersion 
spectra of IW~Per covering different phases were analyzed for these 
lines by using the spectrum-fitting technique to determine the 
abundances of Ca, Sr, and Sc and the corresponding equivalent widths.
It turned out, however, that no firm evidence of such phase-dependent 
line-strength variations could be found, suggesting that significant 
chemical inhomogeneity on the surface of IW Per is unlikely to exist, 
at least as regards to the period of our observations (2010 December).
Meanwhile, the abundances of O, Si, Ca, Ba, and Fe resulting from 
the 6130--6180~\AA\ region corroborate that IW~Per is a distinct Am 
star (though the degree of peculiarity differs from element to element) 
despite that its rotational velocity ($\sim 100$~km~s$^{-1}$) 
is near to the existent limit of Am phenomenon. 
}
{stars: abundances -- stars: chemically peculiar -- stars: early-type -- 
stars: individual (IW Per)  -- stars: spectroscopic binaries}

\section{Introduction}

IW Per is a spectroscopic binary of a short period ($\sim 0.92$~d),  
where lines of only the primary mid A-type star is visible in the spectrum.
It is also a photometric variable with small amplitudes of several 
hundredths mag, which is because of the aspect effect on an ellipsoidal 
stellar shape caused by the considerable centrifugal force due to rapid 
rotation (ellipsoidal variable). 
About four decades ago, Kim (1980) carried out an intensive photometric 
and spectroscopic study of this single-lined binary system, and established
the orbital elements and stellar physical parameters by analyzing the observed 
radial velocity curve as well as the light curve, as summarized in Table~1.    

As an important feature from the viewpoint of spectral classification, this star
belongs to the group of ``A-type metallic-line (Am) stars'', which is a subclass 
of chemically peculiar (CP) stars (upper main-sequence stars showing unusual 
spectra).  Am stars are known to have the following characteristics (see, e.g., 
Conti 1970 or Preston 1974).
\begin{itemize}
\item
Metallic lines (such as Fe) are stronger while the Ca~{\sc ii} 3934 line 
(K line) is weaker in comparison with normal stars, which are used 
(along with Balmer lines) for spectral classification of A-type stars. 
Accordingly, spectral types of Am stars estimated from different indicators 
are discordant with each other as Sp(Ca~K) $<$ Sp(Balmer) $<$ Sp(metal).   
\item
According to chemical abundance studies based on high-dispersion spectra,  
Ca and Sc (as well as light elements such as C, N, and O) are deficient
while heavier elements (Fe group or s-process elements such as Sr or Ba) 
are overabundant in Am stars.
\item
Rotational velocities of Am stars are generally lower (typical $v_{\rm e}\sin i$
is on the order of several tens km~s$^{-1}$), despite that 
many ordinary A stars are rapid rotators (mean $\langle v_{\rm e}\sin i \rangle$
is $\sim$~100--200~km~s$^{-1}$; e.g., Abt \& Morrell 1995). As a matter of fact, 
no Am stars are ever found at $v_{\rm e}\sin i \gtrsim 100$~km~s$^{-1}$. 
\item
There is a trend that Am phenomenon appears especially in spectroscopic 
binaries. This high frequency of Am stars in close binaries may be related 
to the preference of slow rotational velocity mentioned above, because rotation 
tends to be decelerated by tidal braking. 
\end{itemize}

Interestingly, Kim (1980) found from  his spectrograms (used for measuring 
radial velocities) that the strengths of Am-specific lines in IW Per show 
appreciable cyclic variations (by $\sim$~10--20\%) with the rotation phase: 
that is, the equivalent widths of Ca~{\sc ii} 3934 as well as Sc~{\sc ii} 4320 
lines reach the weak bottom around phases of $\sim 0.25$ and $\sim 0.75$, where 
that of Sr~{\sc ii} 4215 line conversely attains the strong peak 
(cf. Fig.~8 of Kim 1980). This implies that the chemical anomaly is not 
uniform over the surface of IW~Per but has rather inhomogeneous or patchy
distribution. 

This is a noteworthy result, because very few Am stars are known to 
exhibit such a phase-dependent line-strength variability, although 
aspect-dependent spectrum variations due to patchy distribution 
of specific elements are actually observed in other class of CP stars 
(e.g., magnetic stars of SrEuCr type). Is IW~Per the first Am star with
inhomogeneous chemical distribution on the surface?

However, very little attention seems to have been paid to this finding so far,
unfortunately. Considering that Kim's result was based on photographic 
spectrograms, which are outdated from the viewpoint of present-day standard, 
it needs to be reconfirmed by all means. 

Accordingly, the author decided to check whether the strengths of Ca~{\sc ii} 3934, 
Sr~{\sc ii} 4215, and Sc~{\sc ii} 4320 lines in IW~Per show such significant 
phase-dependent variations as found by Kim (1980), while using new 
observational materials (10 CCD spectra reasonably covering the phases)
and efficient analysis techniques (spectrum-fitting analysis).
This is the primary purpose of the investigation.
In connection with this inspection, since IW~Per is an ellipsoidal variable with 
inhomogeneous distribution of $T$ (temperature) and $g$ (surface gravity) on 
the distorted surface, the impact of this effect on spectrum variation is 
also estimated based on the Roche model. 

Besides, the quantitative nature of Am anomaly in IW~Per is examined 
by determining the abundances of key elements (O, Si, Ca, Ba, and Fe) 
and comparing them with those of many A-type stars, which is counted
as another aim of the present investigation. This is worthwhile 
because IW~Per is a rapid rotator ($v_{\rm e} \sim 100$~km~s$^{-1}$) 
almost near to the existent limit of Am phenomenon, 

\section{Observational data}

The observations of IW~Per were carried out on 2010 December 14, 
15, 16, 18, and 20 by using BOES (Bohyunsan Observatory Echelle 
Spectrograph) attached to the 1.8 m reflector at Bohyunsan Optical 
Astronomy Observatory in the Republic of Korea. Using 2k$\times$4k CCD 
(pixel size of 15~$\mu$m~$\times$~15~$\mu$m), this echelle spectrograph 
enabled us to obtain spectra of wide wavelength coverage 
(from $\sim$~3800~\AA\ to $\sim$~9200~\AA) with the resolving 
power of $R \simeq 45000$ (use of 200$\mu$m fiber).
The observations were done 1--3 times in a night with an interval 
of a few hours. Thus, as a result of 5-night observations,
10 spectra were obtained in total (each consisting of 1 or 2 
successive frames of 15--20 min exposure to be co-added). 
The total integration time for each spectrum was 15--40 min.
The fundamental information (spectrum code, observed time in 
Julian day, mean S/N ratios around the relevant regions, etc.) 
for each of the 10 spectra is presented in Table 2.

The reduction of the echelle spectra (bias subtraction, flat 
fielding, spectrum extraction, wavelength calibration, and 
continuum normalization) was carried out by using the software package 
IRAF (Image Reduction and Analysis Facility).\footnote{
IRAF is distributed by the National Optical Astronomy Observatories,
which is operated by the Association of Universities for Research
in Astronomy, Inc. under cooperative agreement with the National 
Science Foundation.}

The orbital phase corresponding to the observed time 
(at the mid exposure) of each spectrum was calculated by the 
ephemeris
\begin{equation}
{\rm Minimum} = {\rm JD} 2433617.317 + 0.9171877 E  \;\;\; (E: {\rm integer}) 
\end{equation}
which was taken from Eq.(2) of Kim (1980). 
The observed heliocentric radial velocities ($V_{\rm r}^{\rm hel}$) were 
determined from the spectrum fitting in the 6130--6180~\AA\ region 
(cf. Sect.~4), which are also given in Table~2.  
The resulting $V_{\rm r}^{\rm hel}$ values are plotted against the phase 
in Fig.~1, where the velocity results derived by Kim (1980) are also 
overplotted for comparison. As seen from this figure, both are reasonably 
consistent with each other. 

\section{Standard model atmosphere}

As mentioned in Sect.~1, IW~Per is an ellipsoidal variable, in which
physical conditions on the surface are not uniform, and thus can not 
be described in the strict sense by only one model atmosphere. 
Still, it is meaningful to define the ``standard'' model atmosphere
well representing the averaged physical properties of this star, 
which is to be used for the analysis of the spectra in Sect.~4.
 
The atmospheric parameters for this standard model were established 
in the same manner as done in Takeda et al. (2008, 2009).  
The effective temperature ($T_{\rm eff}$) and surface gravity ($\log g$) 
were evaluated from the colors of Str\"{o}mgren's $uvby\beta$ 
photometric system with the help of the UVBYBETANEW program 
(Napiwotzki et al. 1993). The observed $uvby\beta$ data 
for IW~Per were taken from Paunzen's (2015) compilation. 
Since these colors fit in two processing groups (groups 5 and 6) 
of the UVBYBETANEW program, both groups were attempted, from which 
($T_{\rm eff}$, $\log g$) of (8439~K, 4.21) and (8303~K, 4.22) were 
derived for group 5 and group 6, respectively, Consequently, the 
simple mean of these two results (8371~K, 4.22) were adopted.

Regarding the microturbulence ($v_{\rm t}$), Takeda et al.'s (2008) 
empirical relation (reasonably representing the observed trend  
of $v_{\rm t}$ with uncertainties of $\pm 30\%$) 
\begin{equation}
v_{\rm t} = 4.0 \exp\{- [\log (T_{\rm eff}/8000)/A]^{2}\} \\
\end{equation}
(where $A \equiv [\log (10000/8000)]/\sqrt{\ln 2}$) was adopted,
which gives $v_{\rm t} = 3.89$~km~s$^{-1}$ for $T_{\rm eff} = 8371$~K.

The adopted color data along with the resulting values of $T_{\rm eff}$, 
$\log g$, and $v_{\rm t}$ are also summarized in Table 1.
The model atmosphere  was then constructed by two-dimensionally interpolating 
Kurucz's (1993) ATLAS9 model grid in terms of $T_{\rm eff}$ and $\log g$, 
where the solar-metallicity models were employed.

\section{Spectrum fitting analysis}

The procedures for deriving the abundances and equivalent widths are 
the same as adopted in previous papers (e.g., Takeda et al. 2008, 2009),
which are based on the spectrum fitting program MPFIT developed 
by Takeda (1995).
This program establish the solutions for (i) the abundances of relevant 
elements ($A_{1}, A_{2}, \ldots$), (ii) projected rotational velocity 
($v_{\rm e}\sin i$), and (iii) radial velocity ($V_{\rm r}$) 
by accomplishing the best fit (minimizing $O-C$ residuals) 
between theoretical and observed spectra. 

This fitting analysis was applied to the following four regions in this study.
\begin{itemize}
\item
3925--3942~\AA\ region (Ca and Fe abundances varied)\\
The aim is is to evaluate the equivalent width of Ca~{\sc ii} 3934 line from
the resulting Ca abundance.
\item 
4207--4224~\AA\ region (Sr and Fe abundances varied)\\
The aim is to evaluate the equivalent width of Sr~{\sc ii} 4215 line from
the resulting Sr abundance.
\item 
4316.5--4327.5~\AA\ region (Sc, Ti, and Fe abundances varied)\\
The aim is to evaluate the equivalent width of Sc~{\sc ii} 4320 line from
the resulting Sc abundance.
\item 
6130--6180~\AA\ region (O, Si, Ca, Ba, and Fe abundances varied)\\
The aim is to determine the abundances of O, Si, Ca, Ba, and Fe 
(along with $v_{\rm e}\sin i$ and $V_{\rm r}$)
\end{itemize}
The abundances of all other elements (except for those varied) were 
fixed at the solar abundances during the iterative fitting procedure.

The assumption of LTE (Local Thermodynamic Equilibrium) was applied
to theoretical calculations of all lines in this study.  
The atomic data of the spectral lines were adopted from Kurucz \& Bell's (1995) 
compilation, except for those of 3925--3942~\AA\ region which were taken 
from the VALD database (Ryabchikova et al. 2015). In case that data of damping 
parameters are not available, the default treatment of the WIDTH9 program 
(Kurucz 1993) was employed. The $\lambda$, $\chi_{\rm low}$, and $gf$ values 
of selected important lines are summarized in Table~3.
Figs.~2--5 display how the theoretical spectra for the converged solutions 
fit well with the observed spectra for each region.

Then, after the abundances ($A$) have been established, the equivalent widths 
of Ca~{\sc ii} 3934, Sr~{\sc ii} 4215, and Sc~{\sc ii} 4320 lines ($W_{3934}$, 
$W_{4215}$, and $W_{4320}$) were calculated inversely from $A$(Ca), $A$(Sr), 
and $A$(Sc) with the help of Kurucz's (1993) WIDTH9 program. The final results 
are summarized in Table~4. In Figs.~2--5  are also plotted the resulting $A$ 
as well as $W$ against the corresponding phase.

\section{Model star simulation}

Before discussing the behaviors of Ca~{\sc ii}, Sr~{\sc ii}, and Sc~{\sc ii} lines
with the phase, it is meaningful to check how the aspect change to this ellipsoidal 
variable (where surface physical parameters are position-dependent) potentially 
affects the observed quantities.
For this purpose, the surface structure of the primary star was simulated
based on the Roche model. By using the masses ($M_{1}$, $M_{2}$) and the
distances to the gravity center ($a_{1}$, $a_{2}$) determined by Kim (1980;
cf. Table~1), the shape of the primary $r (\theta, \phi)$ can be
calculated from the equi-potential surface with the boundary condition
of $r (\theta= 0^{\circ}) = 1.7 R_{\odot}$ ($\equiv  R_{\rm p}$: polar radius),
and the distribution of local surface gravity $g(\theta,\phi)$  
($\equiv |{\bf g}_{1} + {\bf g}_{2} + {\bf g}_{\rm cf}|$) is also derived, 
where ${\bf g}_{1}$, ${\bf g}_{2}$, and ${\bf g}_{\rm cf}$ are the gravity force
from $M_{1}$, gravity force from $M_{2}$, and the centrifugal force, respectively.
Further, assuming $T_{\rm eff,p} = 8500~K$ (polar $T_{\rm eff}$ at 
$\theta= 0^{\circ}$) and applying the relation of $T_{\rm eff} \propto g^{1/4}$ 
(von Zeipel's law, which holds for the present case of $T_{\rm eff} \gtrsim 8000$~K),  
$T_{\rm eff}$ is expressed as a function of $(\theta, \phi)$.
The resulting distributions of $r$, $g$, and $T_{\rm eff}$ are illustrated
in Figs.~6a--6c. Comparing these results with those obtained by Kim (1980; 
cf. Sect.~6 in his paper), we can confirm a reasonable consistency with each other.   
  
This distorted rotating star is viewed by an observer with the inclination angle
(angle between the rotational axis and the line of sight) of $i = 63^{\circ}$, 
and the appearance as well as the properties 
of the visible disk change with the rotational phase (cf. Fig.~7a). Here, we define  
$D$ (disk area), $\langle T_{\rm eff} \rangle$ (mean $T_{\rm eff}$ averaged over 
the disk; in K), $\langle \log g \rangle$ (mean $\log g$ averaged over the disk; 
in dex), $F_{5550}$ (observed monochromatic flux by the earth's observer at the 
representative $V$-band wavelength of 5550~\AA\ ; in erg~s$^{-1}$cm$^{-2}$\AA$^{-1}$) as follows.
\begin{equation}
D \equiv \int\!\!\!\int_{{\rm disk}} \; {\rm d}\xi {\rm d}\eta,
\end{equation}
\begin{equation}
\langle \log g \rangle  \equiv 
\left. \int\!\!\!\int_{{\rm disk}} \; \log g(\xi,\eta) \; I_{5550}(\xi,\eta) \; {\rm d}\xi {\rm d}\eta \middle /
\int\!\!\!\int_{{\rm disk}}\; I_{5550}(\xi,\eta) \; {\rm d}\xi {\rm d}\eta \right. ,
\end{equation}
\begin{equation}
\langle T_{\rm eff} \rangle  \equiv 
\left. \int\!\!\!\int_{{\rm disk}} \; T_{\rm eff}(\xi,\eta) \; I_{5550}(\xi,\eta) \; {\rm d}\xi {\rm d}\eta \middle /
\int\!\!\!\int_{{\rm disk}} \; I_{5550}(\xi,\eta) \; {\rm d}\xi {\rm d}\eta \right. ,
\end{equation}
and
\begin{equation}
F_{5550}  \equiv 
\left. \int\!\!\!\int_{{\rm disk}} \; I_{5550}(\xi,\eta) \; {\rm d}\xi {\rm d}\eta \middle / d^{2} \right. ,
\end{equation}
where $I_{5550}(\xi,\eta)$ is the specific intensity at 5550~\AA\ emergent from 
a point on the disk $(\xi, \eta)$ toward the direction of the observer, and
$d$ (= 56.47~pc = $1.742\times 10^{20}$~cm) is the distance to IW~Per (Gaia DR2 parallax is 17.71 milliarcsec).
$I_{5550}$ at each point of the stellar disk was evaluated by interpolating 
the specific intensity grid $I_{\lambda}(\mu, T_{\rm eff}, \log g)$ ($\mu$ is 
the direction cosine) of ATLAS solar metallicity models, which was downloaded 
from the Kurucz site\footnote{http://kurucz.harvard.edu/grids/gridp00/} (filename: ip00k2.pck19). 
The runs of $D$, $\langle \log g \rangle$, $\langle T_{\rm eff} \rangle$, and $F_{5550}$
with the rotation phase are depicted in Figs.~7b--7e.

In Fig~7e, the observed flux of $-0.4V -8.403$ ($V$ is the $V$-band magnitude of 
IW~Per taken from Kim 1980) is also plotted against the phase. Here, the constant of 
$-8.403$ was so chosen (by eye inspection) as to accomplish a match with $\log F_{5550}$. 
Since the absolute calibration relation between $F_{5556}$ (erg~cm$^{-2}$s$^{-1}$\AA$^{-1}$) 
and $V$ (mag) gives $-8.449$ for this constant according to Eq.(10.4) of Gray (2005),
the difference between these two makes $(-8.403)-(-8.449)=0.046$~dex, which may be further 
reduced down to $0.046-0.032 = 0.014$~dex by subtracting 0.032 ($=0.4\times 0.08$)~dex due to 
the interstellar extinction effect of $A_{V} = 0.08 (\pm 0.23)$~mag (estimated by the EXTINCT 
program of Hakkila et al. 1997). This consistency in the offset constant indicates that our 
model reproduces the observed energy distribution of IW~Per satisfactorily (not only the 
relative change with the phase but also in terms of the absolute scale).\footnote{
It should be noted, however, that $F_{5550}$ was calculated based on the disk-integration 
of the model star for the primary, while neglecting the contribution from the secondary, 
which Kim (1980) presumed to be a G0~V star from its mass of $1.08 M_{\odot}$. 
Therefore, this consistency in the absolute sense should not be taken too seriously, 
as it might be something like fortuitous.} 

\section{Search for spectrum variability}

\subsection{Comparison with Kim's (1980) equivalent widths}

We are now ready to discuss whether IW~Per shows any variability in the strengths of 
specific lines characterizing the Am anomaly (Ca~{\sc ii} 3934, Sr~{\sc ii} 4215,
and Sc~{\sc ii} 4320) as reported by Kim (1980). Several characteristic trends can be
seen from Figs.~2c, 3c, and 4c, where his $W$ vs. phase relations (solid 
lines) with those derived in this study (symbols) are compared. 

Kim's (1980) $W$ values tend to be systematically larger especially
for $W_{4215}$ (by $\sim$~0.1--0.2~dex) and $W_{4320}$ ($\sim$~0.1~dex), which 
is likely to be related to the adopted procedure of $W$ evaluation.
Kim (1980) measured $W$ directly from the spectrum with respect to the
empirically placed continuum level. Though this is a conventional approach,
uncertainties are inevitable because target lines are by no means isolated but 
more or less blended with other lines especially in the crowded blue region. 
In contrast, the equivalent width derived in this study represents only the pure
contribution of the line in question, because it was calculated from the 
abundance of the relevant element resulting from the spectrum fitting analysis.
It is thus understandable that such a methodological difference may have caused 
an appreciable discrepancy in the absolute values of $W$. 

Therefore, we focus on the nature of ``relative variation'' in 
$\log W$. Are the phase-dependent strength variations of Am-specific 
lines reported by Kim (1980) (i.e., Am anomalies are enhanced 
around phases of $\sim 0.25/0.75$ while weakened around $\sim 0.0/0.5$) 
confirmed also in our observational data? \\
--- Interestingly, somewhat similar trend may be seen for the Ca~{\sc ii} 3934 
line, since it appears to show a weak variation of $W_{4215}$ with a minimum 
at phase of $\sim 0.8$ (Fig.~2c) like the case of Kim (1980).\\ 
--- However, regarding the other two lines (Sr~{\sc ii} 4215 
and Sc~{\sc ii} 4320), we do not see any signature of meaningful cyclic variation 
in their strengths (Figs.~3c and 4c) unlike his measurements. \\
--- Especially, only small fluctuation of $W_{4215}$ by a few percent 
($\pm \lesssim 0.02$~dex) seriously conflicts Kim's (1980) result not only in 
terms of its size (the change he found amounts to $\sim \pm 0.05$~dex) but also 
in the sense of variation (our $W_{4215}$ shows a minimum around phase $\sim 0.2$ 
where his $W_{4215}$ has a maximum).\\
--- Accordingly, it is reasonable to state that our observational data
could not confirm the previous results of Kim (1980).

\subsection{Origin of line strengths fluctuations}

Here, it is worth discussing the cause of fluctuations in our $W$ 
($\pm \sigma_{W}/\langle W \rangle$), which are $\sim \pm 0.05$, $\sim \pm 0.04$, and 
$\sim \pm 0.03$ according to Table~4 for Ca~{\sc ii} 3934, Sr~{\sc ii} 4215,
and Sc~{\sc ii} 4320, respectively (or $\sim \pm 0.02$~dex, $\sim \pm 0.017$~dex,
and $\sim \pm 0.01$~dex in $\log W$).

First, measurement errors involved in our $W$ values should be examined.
Here, the most important factor is the uncertainty in the continuum level.\footnote{
Errors in $W$ due to pure photometric origin ($\delta W$) were also estimated 
by using Cayrel's (1988) formula (depending on S/N, pixel size, and line widths) but 
turned out insignificant:  $\delta W_{3934} \sim$~2--5~m\AA\ ($\sim 10$~m\AA\ only for 1220C), 
$\delta W_{4215} \sim$~1--3~m\AA\, and $\delta W_{4320} \sim$~1--2~m\AA.} 
The spectrum-fitting method (Takeda 1995) adopted in Sect.~4 does not require
any empirical specification of the continuum position ($F_{\rm c}$) in advance; 
but $F_{\rm c}$ is theoretically determined after the fitting has been established.
Any kind of spectrum change (e.g., due to noise) would cause an ambiguity in $F_{\rm c}$.
The equivalent width is approximately expressed as $W \simeq H (1 - F_{0}/F_{\rm c})$, 
where $F_{0}$ is the line-center flux and $H$ is the full-width at half-maximum of the line.
If the continuum level is perturbed as $F_{\rm c}' = F_{\rm c}(1 + \epsilon)$
(where $\epsilon$ is the relative error in $F_{\rm c}$; $|\epsilon| \ll 1$) 
and $H$ as well as $F_{0}$ are assumed to be unchanged, then the ratio of the 
corresponding equivalent widths ($W'$ and $W$) can be written as 
\begin{equation}
W'/W \simeq 1 + \epsilon (F_{0}/F_{\rm c})/(1 - F_{0}/F_{\rm c}).
\end{equation}  
Since this relation suggests $W'/W \sim 1 + \epsilon /(1 - F_{0}/F_{\rm c})$
at the weak-line limit ($F_{0}/F_{\rm c} \lesssim 1$)
and  $W'/W \simeq 1 + \epsilon (F_{0}/F_{\rm c}) (\simeq 1) $ (i.e., 
practically $\epsilon$-independent) at the strong-line limit
($F_{0}/F_{\rm c} \simeq 0$), the effect of $\epsilon$ on $W'/W$ becomes 
progressively important as the line gets weaker.
Let us tentatively set $\epsilon (\sim ({\rm S/N})^{-1}) = \pm 0.01$ 
(for Ca~{\sc ii} 3934) and $\pm 0.005$ (for Sr~{\sc ii} 4215 and Sc~{\sc ii} 4320) 
based on the typical S/N ratios given in Table~2.
Considering that the relevant $F_{0}/F_{\rm c}$ values are $\sim 0.2$, 
$\sim 0.85$, and $\sim 0.9$ (relative to the local H$\gamma$ wing) 
for Ca~{\sc ii} 3934, Sr~{\sc ii} 4215, and Sc~{\sc ii} 4230, 
the resulting changes of $\log (W'/W)$ are $\pm 0.001$~dex, $\pm 0.01$~dex, 
and $\pm 0.02$~dex, respectively.

Next, we estimate the effects due to the modulation of mean atmospheric 
parameters, which are $\sim \pm 0.005$~dex in $\langle \log g \rangle$ (Fig.~7c) 
and $\sim \pm 20$~K in $\langle T_{\rm eff} \rangle$ (Fig.~7d).
According to the parameter sensitivity of $W_{3934}$, $W_{4215}$, and $W_{4320}$ 
shown in Table~5, the former $\log g$ effect is essentially negligible ($< 0.001$~dex
in any case), while the latter $T_{\rm eff}$ effect (relative change in $W$ for 
$\pm 20$~K) is $\pm 0.018$, $\pm 0.006$, and $\pm 0.008$ (or $\pm 0.0077$~dex,
$\pm 0.0026$~dex, and $\pm 0.0035$~dex in $\log W$). It is worth keeping in mind that
the Ca~{\sc ii} 3934 line is more sensitive to this $\langle T_{\rm eff} \rangle$
modulation (by a factor of $\sim$~2--3) than the other 2 lines. 

Based on these two rough considerations, it may be possible to explain (at least 
qualitatively) the cause of observed fluctuations (by $\sim \pm$~0.01--0.02~dex)
in our $\log W_{3934}$, $\log W_{4215}$, and $\log W_{4320}$.\\
--- Regarding the strong Ca~{\sc ii} 3934 line, while the effect of continuum 
uncertainty is inessential, the cyclic change of $\langle T_{\rm eff}\rangle$
by $\pm 20$~K may cause a variation of $\sim \pm 0.01$~dex in $\log W_{3934}$.
That the minimum of $W_{3934}$ is observed around phase~$\sim 0.8$ (cf. Fig.~2c)
is consistent with this interpretation, because $\langle T_{\rm eff}\rangle$ attains
a maximum (making $W_{3934}$ weaker) around this phase.\\ 
--- In contrast, as to the weaker Sr~{\sc ii} 4215 and Sc~{\sc ii} 4230 lines,
even slight uncertainties (by $\pm$ several tenths percent) in the continuum position 
would result in appreciable changes by $\sim \pm$~0.01--0.02~dex in $\log W_{4215}$ 
as well as $\log W_{4320}$, to which the observed fluctuations may be attributed.

\subsection{Absence of non-uniform surface chemical anomaly}

As described in Sect.~6.1, our $W_{3934}$, $W_{4215}$, and $W_{4320}$ values 
do not show clear cyclic variations, contrary to Kim's (1980) finding.   
It was also shown in Sect.~6.2 that the extents of their fluctuations may be 
understood in terms of possible errors in $W$ due to continuum uncertainty (plus 
the effect of slight change in $\langle T_{\rm eff} \rangle$ for Ca~{\sc ii} 3934).   
Therefore, as far as our observational data are concerned, we do not see any 
evidence for an inhomogeneous distribution of Am anomaly on the surface of IW~Per
suspected by Kim (1980).

Also, this consequence is corroborated by the spectrum-fitting analysis of the 
6130--6180~\AA\ region (Fig.~5a), which yielded the abundances of O, Si, Ca, Ba, and Fe.
Since lines are less crowded in this orange region as compared with the 
blue--violet region, the resulting abundances are considered to be more reliable.
As seen from Figs.~5b--5f, no clear phase-dependence is observed in the resulting 
abundances showing only small dispersions ($\sigma_{A}$ is $\sim$~0.01--0.02~dex for 
O, Ca, Fe: $\sim$~0.04~dex for Si and Ba; cf. Table~4). Moreover, we do not see 
any distinct interrelation between these abundances. This may be counted as
another counter-example, because such an inhomogeneous distribution of Am 
peculiarity (if any exists) would reveal some kind of anticorrelation 
between the deficiency group (O, Ca) and enrichment group (Si, Ba, Fe).    

Although our conclusion is in conflict with that of Kim (1980), this does not
necessarily mean that his result was incorrect. What we can state is that non-uniform 
Am anomaly on the surface of IW~Per is unlikely at the time of our observations 
(2010 December). We can not rule out a possibility that the nature of Am phenomenon 
changes in the course of time and it might have been actually patchy at the period of 
Kim's (1980) spectroscopic observations (1976--1977). Generally speaking, however, 
Kim's (1980) results had better be viewed with caution because his equivalent widths 
were measured on photographic spectrograms of insufficient quality. 

\section{Status of IW Per as an Am star}

Another significant topic related to IW~Per is its degree of Am peculiarity.
Since synchronization is considered to be achieved in this short-period close 
binary system, the equatorial rotational velocity of the primary is expected 
as $v_{\rm e} = (2\pi \times 1.05 R_{\rm p})/P = 99$km~s$^{-1}$, which is 
in agreement with the value of $v_{\rm e} = 98$~km~s$^{-1}$ derived from the 
observed $v_{\rm e}\sin i$ ($=87$~km~s$^{-1}$; Table~4) and known $i$ ($=63^{\circ}$; Table~1).
Therefore, IW~Per is characterized by its high rotational velocity 
($v_{e} \simeq 100$~km~s$^{-1}$) for an Am star, which is near to the upper limit 
for the appearance of Am phenomenon ($v_{\rm e}\sin i \lesssim 100$~km~s$^{-1}$; cf. Sect.~1).

In view of the fact that Am stars are found only in this comparatively lower 
$v_{\rm e}\sin i$ range, it has been a controversial issue whether the degree of 
anomaly progressively fades away with an increase of $v_{\rm e}\sin i$ (e.g., 
Kodaira 1975; Takeda \& Sadakane 1997) or Am peculiarity tends to persist without 
such a systematic change up to the limit (e.g., Burkhart 1979).
It is thus meaningful to examine the nature and extent of chemical anomaly 
in this rare rapidly-rotating Am star by comparing with other normal A and Am 
stars of various rotational velocities. 

Conveniently, Takeda et al. (2009) previously carried out an extensive spectroscopic 
investigation for 122 A-type stars (including 23 Am stars and 23 Hyades stars)\footnote{
By consulting the spectral types in three compilations (SIMBAD database, Bright Star 
Catalogue, and Hipparcos catalogue), those classified as ``Am'' in at least two out 
of these three sources were regarded as Am stars in this study: they are 
 HD~20320,  23281,  27045,  27628,  27749,  28226,  28546,
  29479,  29499,  30121,  30210,  33204,  33254,  33641,
  40932,  48915,  72037,  95608, 141795, 173648, 198639,
 204188, and 207098. Meanwhile, 23 Hyades stars are the same as indicated
in Table~1 of Takeda et al. (2009).} 
covering a wide range of rotational velocities ($10 \lesssim v_{\rm e}\sin i \lesssim 300$~km~s$^{-1}$).
Although the main attention of that paper was paid to the Na~{\sc i} 5890/5896 lines of 
these stars,their abundances of O, Si, Ca, Ba, and Fe were also determined (cf. Table~1 in 
Takeda et al. 2009) by the spectrum fitting in the 6140--6170~\AA\ region (similar to 
that we have done for IW~Per; cf. Fig.~5), which makes us possible 
to discuss how IW~Per is compared with  other A-type stars in terms of the abundances 
of these Am-specific elements.  

The behaviors of the [X/H] values (abundances relative to the standard star Procyon
which is known to have almost the solar composition, where X is any of O, Si, Ca, Ba, 
and Fe)\footnote{
[X/H]$_{*} \equiv A_{*} - A_{\rm procyon}$, where $A_{\rm procyon}$ is 
8.87 (O), 7.14 (Si), 6.19 (Ca), 2.33 (Ba), and 7.49 (Fe), which were
determined from the 6140--6170~\AA\ fitting in the similar manner (cf. the footnote in 
Sect.~IVc of Takeda et al. 2008). IW~Per have [O/H] = $-0.19$, [Si/H] = $+0.01$, 
[Ca/H] = $-0.29$, [Ba/H] = +0.99, and [Fe/H] = +0.15.} 
for IW~Per and 122 A-type stars are shown in Fig.~8, where their dependences 
upon $T_{\rm eff}$ and $v_{\rm e}\sin i$ as well as mutual correlations are displayed.
An inspection of Fig.~8 reveals the following trends.
\begin{itemize}
\item
It is no doubt that IW~Per is an Am star because the resulting abundances distinctly show
characteristic trends of Am phenomena (e.g., underabundance in O and Ca, overabundance
in Ba and Fe). 
\item
However, answering the question about the degree of Am anomaly in IW~Per is not so 
simple as it appears to differ from element to element. (i) Regarding O and Ca (which generally 
show deficiency in Am stars), the extent of peculiarity in this star is apparently less 
manifest as compared to slowly rotating Am stars (Figs. 8f and 8h). (ii) Meanwhile, 
the abundances of Si and Fe (both are are correlated and tend to be enriched) in IW~Per 
are not much different from those of other Am stars (Figs. 8g and 8j). (iii) Yet, it is 
remarkable that IW~Per shows a conspicuous overabundance of Ba by as large 
as $\sim +1$~dex in spite of its being a rapid rotator (Fig. 8i). 
\item
Therefore, while it is intuitively reasonable that O and Ca (deficiency group) show 
weaker peculiarity in IW~Per (because rapid rotation near to the upper limit of 
Am phenomenon should have acted to suppress the anomaly), it is hard to understand 
why Ba (representative enrichment group) is so anomalously overabundant in the same star.  
\item
These results suggest that the impact of rotation on the chemical anomaly is likely
to act differently on each element in a complex way. Likewise, as to the 
issue of whether or not the abundance peculiarity in Am stars progressively
declines with an increase in $v_{\rm e}\sin i$, it is hard to give a simple answer, 
because some elements exhibit such a systematic trend (e.g., O, Ca) while others do not
(e.g., Ba, showing rather a large dispersion).    
\end{itemize}

\section{Summary and conclusion}

IW Per is a single-lined spectroscopic binary, which shows a slight periodic 
change in its brightness, since the orbital period is so short (0.92~d) 
that the stellar shape is distorted due to the centrifugal force (ellipsoidal 
variable). 

This star is classified as belonging to the group of A-type metallic-line (Am) 
stars, which are often found in slow rotators as well as in close binary 
systems, and generally show anomalous line strengths of specific elements 
(e.g., O, Ca, Sc lines are weak, while lines of Fe-peak elements or s-process 
group are strong). It is noteworthy that IW Per is a rare Am star of rapid 
rotation ($v_{\rm e} \simeq 100$~km~s$^{-1}$) near to the upper limit for 
the appearance of chemical peculiarity.

Previously, Kim (1980) reported based on his spectroscopic
observations of IW~Per that the equivalent widths of Ca~{\sc ii} 3934, 
Sr~{\sc ii} 4215, and Sc~{\sc ii} 4320 lines (important lines characterizing 
the Am anomaly) show cyclic variations in accordance with the rotation phase.
This is an important implication that chemical peculiarities on 
the surface of IW~Per are not uniformly distributed  but are rather patchy  
(an unusual case which is barely found in other Am stars).
Unfortunately, this finding has acquired little attention and any follow-up 
study for its reconfirmation seems to have never been tried so far. 

Motivated by this situation, 10 high-dispersion spectra of IW~Per covering 
different phases were analyzed by applying the spectrum-fitting technique  
to the regions comprising these lines to determine the abundances 
of Ca, Sr, and Sc, from which the relevant equivalent widths 
($W_{3934}$, $W_{4215}$, $W_{4320}$) were calculated. Likewise, in order 
to examine the characteristics of Am anomaly, the abundances of O, Si, Ca, 
Ba, and Fe were also established from the fitting analysis in the 
6130--6180~\AA\ region.

The resulting $W$ data revealed no such clear phase-dependent
variations as found by Kim (1980) (though some partly similar trend 
can not be excluded for $W_{3934}$). It is evident, however, $W_{4215}$ 
as well as $W_{4320}$ are essentially phase-independent and their 
fluctuations are reasonably explained by S/N-related uncertainties 
of the continuum level, which suggests that patchy abundance 
distributions of these elements are improbable.

Accordingly, contrary to the finding of Kim (1980), we conclude that
any appreciable inhomogeneity of abundance anomaly on the surface of 
IW~Per is unlikely, at least at the time of our observations 
in 2010 December. Although this consequence does not necessarily 
mean that Kim's (1980) result was incorrect (i.e., the situation might 
have changed with time), his measurements based on photographic 
spectrograms had better be viewed with caution.

In connection with searching for any variability in $W$, for that 
the local atmospheric parameters ($T_{\rm eff}$ and $\log g$) of an 
ellipsoidal variable such as IW~Per weakly depend upon the position 
on the surface, its impact on the spectrum variation 
of a rotating star was also checked based on a model star of Roche 
potential surface. But this effect on the line strength was found to 
be insignificant (except for the Ca~{\sc ii} 3934 line, which may 
suffer some change by this cause). 

Finally, the abundances of O, Si, Ca, Ba, and Fe confirm that IW~Per 
has typical abundance characteristics of Am stars (O and Ca
are deficient; Ba and Fe are overabundant), despite that its rapid 
rotation is almost near to the upper limit of Am stars.
Yet, it is not easy to answer the question whether the chemical anomaly 
in this rapidly rotating Am star is quantitatively less manifest 
as compared to other slower rotators, because the degree of peculiarity 
differs from element to element (e.g., while deficits in O and Ca are 
surely milder, excess in Ba is still considerably large). 

\Acknow{This investigation has made use of the SIMBAD database, operated by CDS, 
Strasbourg, France, and the VALD database operated at Uppsala University,
the Institute of Astronomy RAS in Moscow, and the University of Vienna.}

\newpage

\begin{figure}[h]
\begin{minipage}{150mm}
\begin{center}
\includegraphics[width=7.0cm]{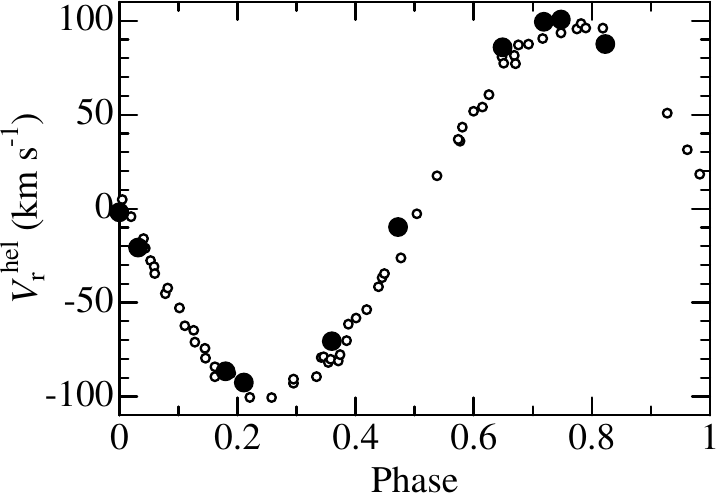}
\end{center}
\FigCap{Observed heliocentric radial velocities of IW~Per (cf. Table~2) 
are plotted against the phase by larger filled symbols, while Kim's (1980) 
results are also overplotted by smaller open symbols for comparison.
}
\end{minipage}
\end{figure}

\begin{figure}[h]
\begin{minipage}{150mm}
\begin{center}
\includegraphics[width=9.0cm]{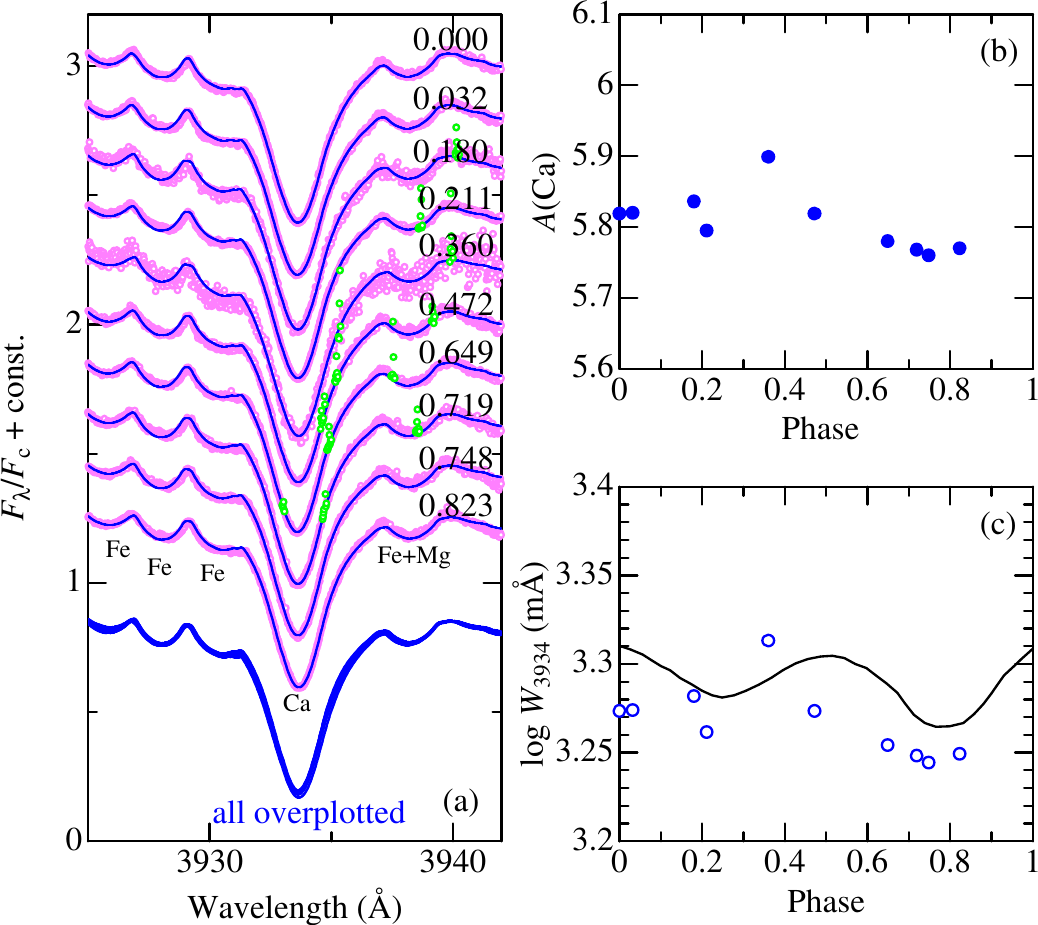}
\end{center}
\FigCap{(a) Synthetic spectrum fitting in the 3925--3942~\AA\ region comprising 
the Ca~{\sc ii} 3934 line. 
The observed 10 spectra at different phases (each shifted vertically relative to the 
adjacent one) are plotted in pink symbols (where the masked pixels are colored in 
light-green) while the best-fit theoretical spectra are shown in blue solid lines. 
The wavelength scale is adjusted to the laboratory system.  
In the bottom of the panel, all of these 10 theoretical spectra are overplotted
(their residual flux scale is in the left ordinate) in order to show
how they are compared with each other. 
(b) Ca abundances derived from this spectrum fitting (cf. Table~4) are plotted against the phase.
(c) Equivalent widths of the Ca~{\sc ii} 3934 line (cf. Table~4) are plotted against the phase, 
where the relation derived by Kim (1980; read from his Fig.~8) is
also plotted by the solid line for comparison.    
}
\end{minipage}
\end{figure}

\begin{figure}[h]
\begin{minipage}{150mm}
\begin{center}
\includegraphics[width=9.0cm]{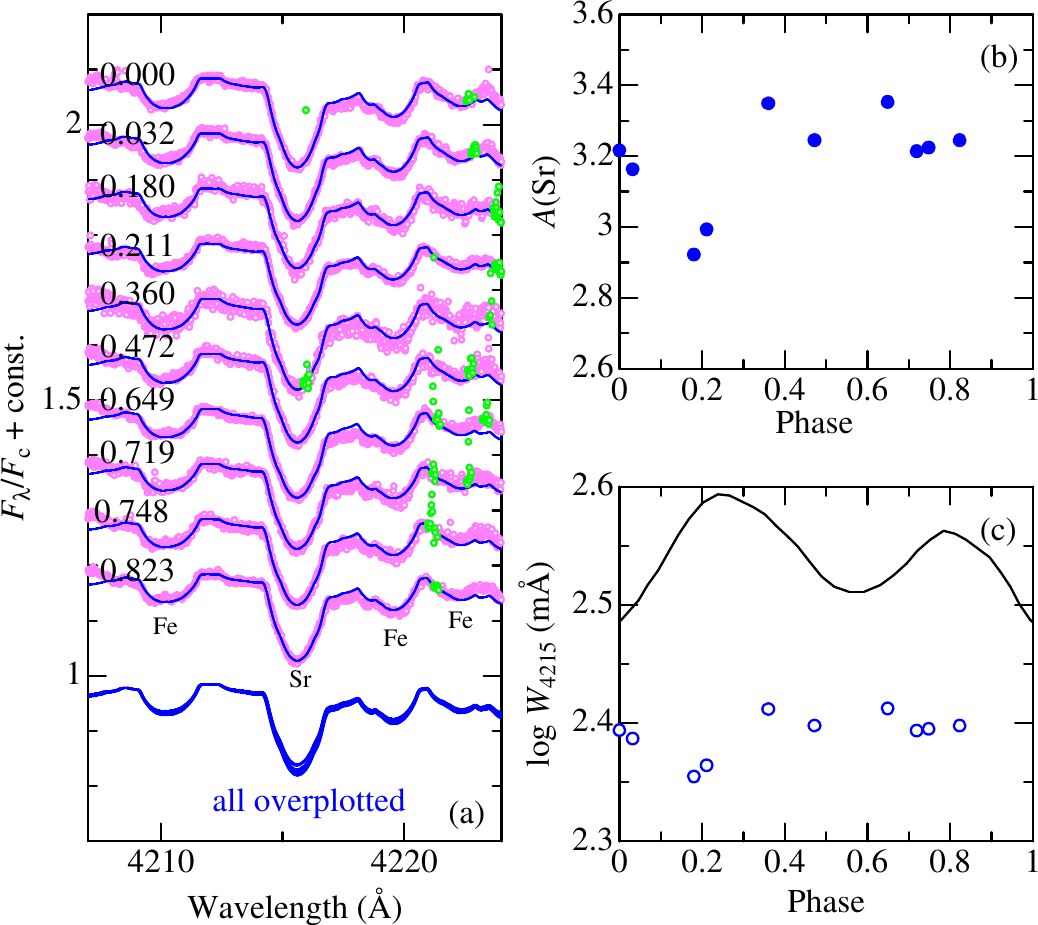}
\end{center}
\FigCap{(a) Synthetic spectrum fitting in the 4207--4224~\AA\ region comprising 
the Sr~{\sc ii} 4215 line. (b) Sr abundance vs. phase relation.
(c) Sr~{\sc ii} 4215 equivalent width vs. phase relation. Otherwise, the same as in Fig.~2.}
\end{minipage}
\end{figure}

\begin{figure}[h]
\begin{minipage}{150mm}
\begin{center}
\includegraphics[width=9.0cm]{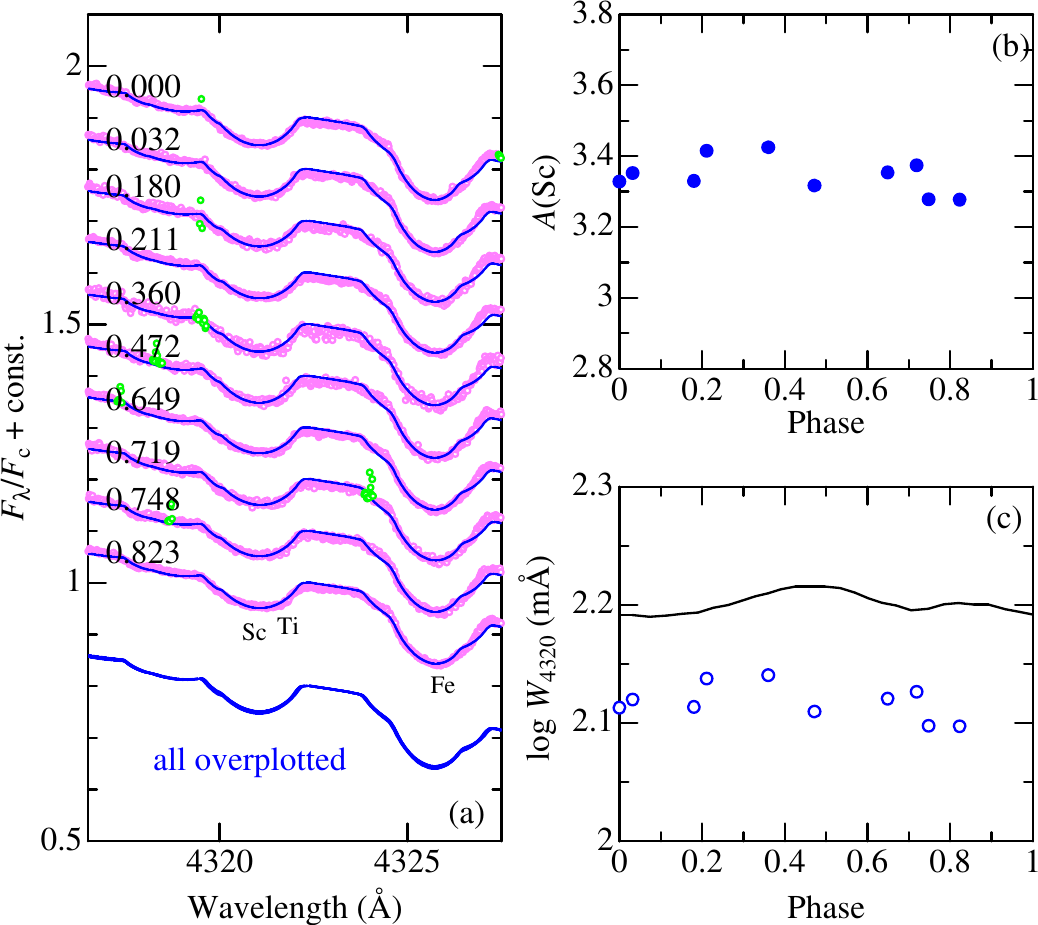}
\end{center}
\FigCap{(a) Synthetic spectrum fitting in the 4316.5--4327.5~\AA\ region comprising 
the Sc~{\sc ii} 4320 line. (b) Sc abundance vs. phase relation.
(c) Sc~{\sc ii} 4320 equivalent width vs. phase relation. Otherwise, the same as in Fig.~2.}
\end{minipage}
\end{figure}

\begin{figure}[h]
\begin{minipage}{150mm}
\begin{center}
\includegraphics[width=9.0cm]{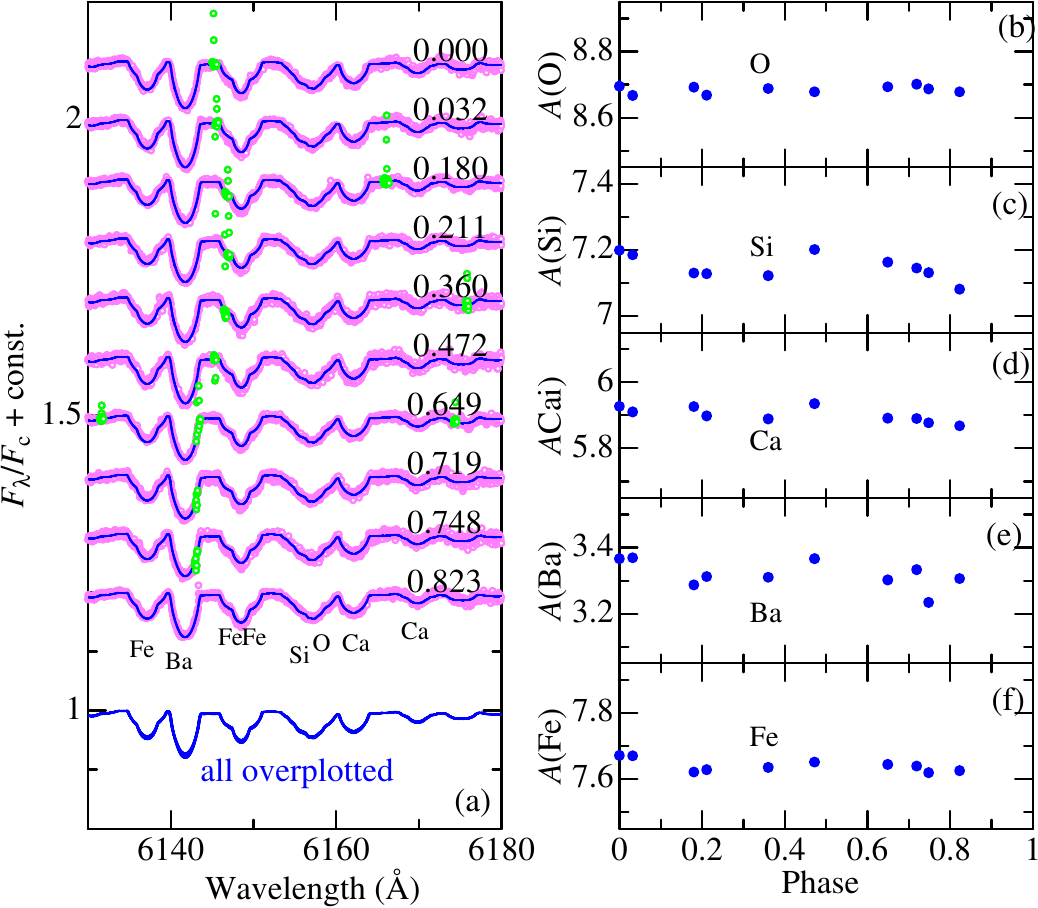}
\end{center}
\FigCap{(a) Synthetic spectrum fitting in the 6130--6180~\AA\ region comprising 
the lines of O, Si, Ca. Ba, and Fe. (b)--(f) The abundances of O, Si, Ca. Ba, 
and Fe derived from the fitting (cf. Table~4) are plotted against the phase. 
Otherwise, the same as in Fig.~2.}
\end{minipage}
\end{figure}

\begin{figure}[h]
\begin{minipage}{150mm}
\begin{center}
\includegraphics[width=10.0cm]{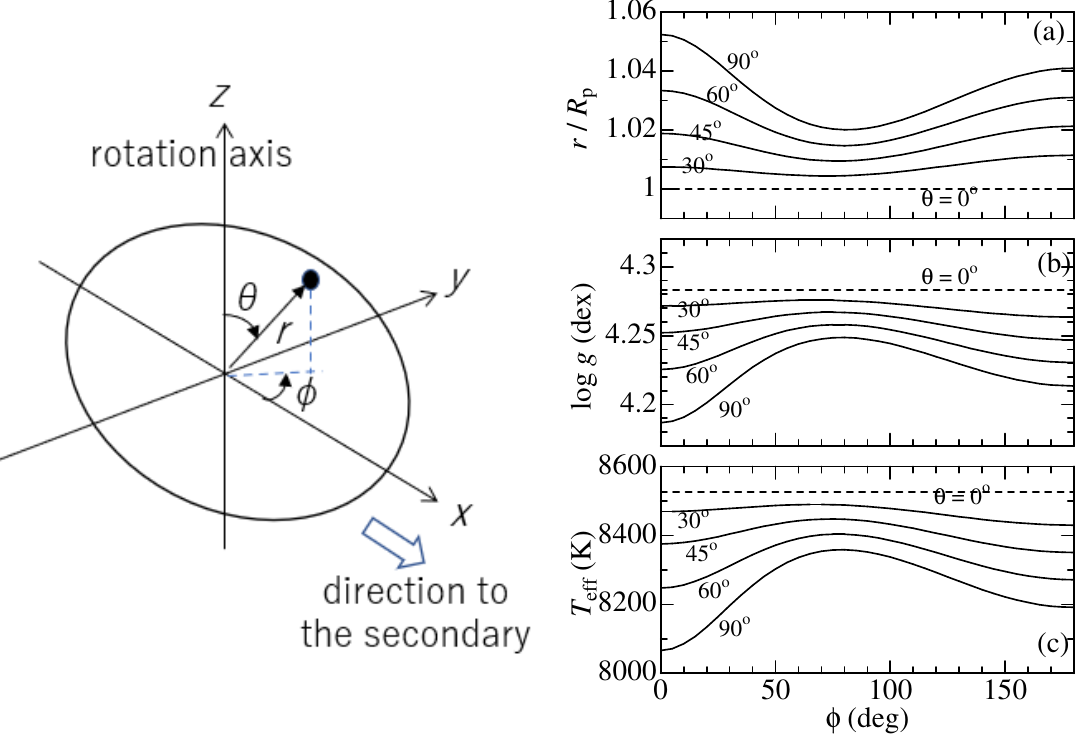}
\end{center}
\FigCap{Roche model calculated for the primary of IW~Per. The parameters 
at the surface are expressed in the spherical coordinate system ($r$, $\theta$, $\phi$; 
see the left-hand side of this figure for their definition).
(a) Radius, (b) surface gravity, and (c) effective temperature.
}
\end{minipage}
\end{figure}

\begin{figure}[h]
\begin{minipage}{150mm}
\begin{center}
\includegraphics[width=11.0cm]{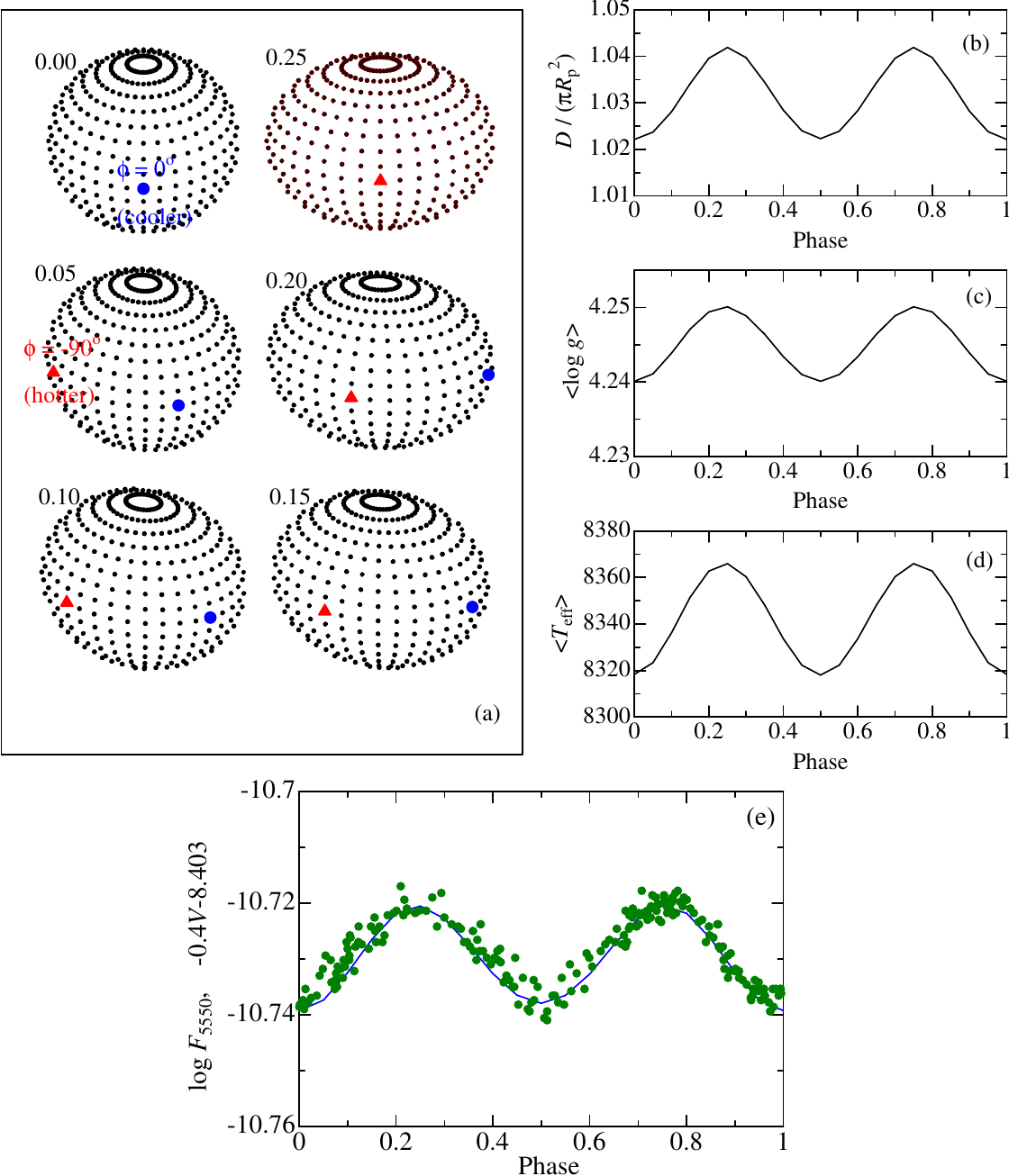}
\end{center}
\FigCap{(a) Schematic view of the primary of IW~Per for six representative phases;
0.00, 0.05, 0.10, 0.15, 0.20, and 0.25, where the oblateness of star shape is purposely exaggerated. 
The $\phi = 0^{\circ}$ and $\phi = -90^{\circ}$ points at the equator 
($\theta = 90^{\circ}$) are also indicated by filled circles and filled triangles, respectively.
(b) Run of the visible disk area ($D$, in unit of $\pi R_{\rm p}^2$) with phase. 
(c) Run of the disk-averaged surface gravity
($\langle \log g \rangle$) with phase. (d) Run of the disk-averaged effective temperature
($\langle T_{\rm eff} \rangle$) with phase. 
(e) The run of $F_{5550}$ (theoretical flux at 5550~\AA\ at the earth in unit of 
erg~cm$^{-2}$s$^{-1}$\AA$^{-1}$, which is calculated by disk integration) with phase is shown by 
the solid line. The observed fluxes corresponding to the $V$ magnitudes of IW~Per 
(derived from the $\Delta V$ data given in Table~IVa of Kim 1980 along with 
his comparison star magnitude of $V = 6.52$) are also depicted by symbols (cf. Sect.~5 
for the explanation on the offset constant of $-8.403$).
}
\end{minipage}
\end{figure}

\begin{figure}[h]
\begin{minipage}{150mm}
\begin{center}
\includegraphics[width=11.0cm]{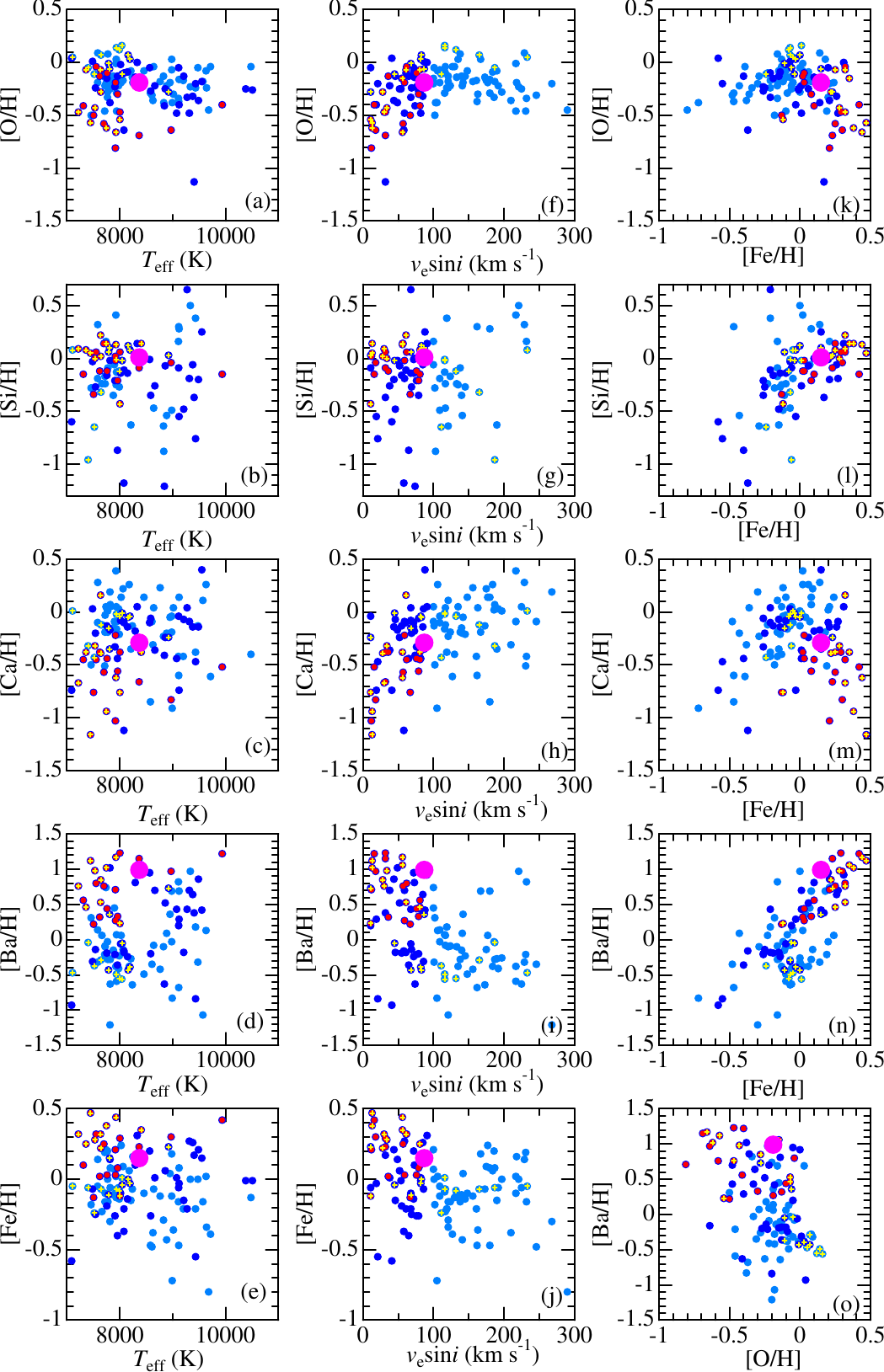}
\end{center}
\FigCap{
Graphical display showing the characteristics of [O/H], [Si/H], [Ca/H], [Ba/H], 
and [Fe/H] derived for IW~Per from the 6130--6180~\AA\ fitting (differential abundances 
of O, Si, Ca, Ba, and Fe relative to the standard star Procyon; practically 
the same composition as the Sun) in comparison with those of 
122 A-type stars previously determined by Takeda et al. (2009),
where the data for IW~Per are highlighted by pink large symbols.   
These [O/H], [Si/H], [Ca/H], [Ba/H], and [Fe/H] values are plotted against $T_{\rm eff}$
in the left panels (a--e) and against $v_{\rm e}\sin i$ in the center panels (f--j).
In the right panels (k--o) are shown the mutual correlations between these abundances: 
[X/H] values (X = O, Si, Ca, Ba) are plotted against [Fe/H] in panels (k--n), 
while the right-bottom panel (o) displays the [Ba/H] vs. [O/H] diagram.  
Regarding 122 A-type stars. those of lower $v_{\rm e}\sin i$ ($ < 100$~km~s$^{-1}$)
and higher $v_{\rm e}\sin i$ ($\ge 100$~km~s$^{-1}$) are discriminated in 
dark blue and light blue, respectively. Also, those 23 stars classified as Am are 
distinguished by overplotting small red-filled circles, while yellow crosses are
overplotted for 23 Hyades cluster stars.
}
\end{minipage}
\end{figure}

\setcounter{table}{0}
\begin{table}[h]
\caption{Orbital and stellar parameters of IW Per adopted in this study}
\scriptsize
\begin{center}
\begin{tabular}{cccl}\hline\hline
Parameter & Value &  Unit & Explanation \\
\hline
\multicolumn{4}{c}{[Orbital and stellar parameters taken from Kim (1980)]} \\
$P$   & 0.9171877  & day  & orbital period \\
$T_{0}$  & 2433617.317  & Julian Day  & Origin of light minimum (phase = 0) \\
$K$   & 99.3  & km~s$^{-1}$  &  Radial velocity amplitude\\
$\gamma$  & 0.2  & km~s$^{-1}$  & system radial velocity \\
$e$   & 0.02  & $\cdots$  & ellipticity\\
$\omega$  & 106.9  & deg  & longitude of periastron\\
$i$   & 63  & deg  & inclination angle \\
$M_{1}$  & 2.0  & $M_{\odot}$  & mass of the primary\\
$M_{2}$  & 1.08  & $M_{\odot}$  & mass of the secondary\\
$R_{1}$  & 1.7  & $R_{\odot}$  & radius of the primary ($\equiv R_{\rm p}$; polar radius)\\
$R_{2}$  & 1.0  & $R_{\odot}$  & radius of the secondary\\
$a_{1}$  & 2.03  & $R_{\odot}$  & distance between the primary and the center of gravity\\
$a_{2}$  & 3.76  & $R_{\odot}$  & distance between the secondary and the center of gravity\\
$s_{12}$ & 5.79  & $R_{\odot}$  & separation between the primary and the secondary\\
\hline
\multicolumn{4}{c}{[Colors and atmospheric parameters]} \\
$^{*}b-y$ & 0.057  & mag  & Str\"{o}mgren's $b-y$ color \\
$^{*}m_{1}$  & 0.229  & mag  & Str\"{o}mgren's $m_{1}$ index\\
$^{*}c_{1}$  & 0.945  & mag  & Str\"{o}mgren's $c_{1}$ index\\
$^{*}\beta$ & 2.880  & mag  & Str\"{o}mgren's $\beta$ index\\
$^{\dagger}T_{\rm eff}$ & 8371  & K  & effective temperature\\
$^{\dagger}\log g$ & 4.22  & dex  & logarithmic surface gravity (in c.g.s.)\\
$^{\#}v_{\rm t}$  & 3.89  & km~s$^{-1}$  & microturbulence\\
\hline
\end{tabular}
\end{center}
$^{*}$Taken from Paunzen (2015).\\
$^{\dagger}$ Derived from colors of the Str\"{o}mgren system by using Napiwotzki et al.'s (1993) 
UVBYBETANEW program.\\
$^{\#}$Derived from Takeda et al.'s (2008) $T_{\rm eff}$-dependent empirical relation\\ 
\end{table}

\setcounter{table}{1}
\begin{table}[h]
\caption{Basic information of the observational spectra}
\scriptsize
\begin{center}
\begin{tabular}{cccccrcccc}\hline\hline
Sp.code & $N_{\rm f}$ & Exp.T&   HJD   &  Phase & $V_{\rm r}^{\rm hel}$ & 
SN$_{3934}$ & SN$_{4215}$ & SN$_{4320}$ & SN$_{6150}$ \\
(1) & (2) & (3) & (4) & (5) & (6) & (7) & (8) & (9) & (10) \\
\hline
1218B &  2 & 35 &  2455549.110 &  0.000 &  -2.0 & 130 &  300 & 300 &  350 \\
1220A &  1 & 15 &  2455550.973 &  0.032 & -20.7 & 170 &  280 & 290 &  350 \\
1218C &  1 & 20 &  2455549.274 &  0.180 & -86.6 &  80 &  170 & 230 &  350 \\
1220B &  2 & 30 &  2455551.138 &  0.211 & -92.6 & 160 &  330 & 370 &  360 \\
1220C &  1 & 20 &  2455551.274 &  0.360 & -70.6 &  30 &  110 & 140 &  250 \\
1214A &  2 & 40 &  2455544.956 &  0.472 &  -9.9 & 100 &  210 & 230 &  240 \\
1216A &  2 & 30 &  2455546.953 &  0.649 & +85.8 & 160 &  280 & 340 &  380 \\
1214B &  2 & 40 &  2455545.183 &  0.719 & +99.4 & 120 &  220 & 260 &  280 \\
1215A &  2 & 40 &  2455546.126 &  0.748 &+100.6 & 130 &  220 & 260 &  290 \\
1218A &  2 & 30 &  2455548.947 &  0.823 & +87.6 & 200 &  340 & 430 &  390 \\
\hline
\end{tabular}
\end{center}
(1) Spectrum code. For example, ``1218B'' denotes the second [B] observation 
on 2010 December 18 [1218], or ``1216A'' corresponds to the first [A] observation
on 2010 December 16 [1216]. (2) Number of frames co-added to obtain the final spectrum.
(3) Total exposure time (in unit of min). (4) Heliocentric Julian Day at the time of 
mid-exposure. (5) Orbital phase calculated by Eq.~(1). (6) Heliocentric
radial velocity (in unit of km~s$^{-1}$) determined by the spectrum fitting 
in the 6130--6180~\AA\ region. (7) Mean S/N ratio around the Ca~{\sc ii}~3934 line.
(8) Mean S/N ratio around the Sr~{\sc ii}~4215 line. (9) Mean S/N ratio around
the Sc~{\sc ii}~4320 line. (10) Mean S/N ratio in the 6130--6180~\AA\ region. \\
Note that these data are arranged in the ascending order of the orbital phase.
\end{table}

\setcounter{table}{2}
\begin{table}[h]
\caption{Atomic data of important lines.}
\tiny
\begin{center}
\begin{tabular}{cccccccc}\hline\hline
Species & $\lambda$ & $\chi_{\rm low}$ & $\log gf$ & Species & $\lambda$ & $\chi_{\rm low}$ & $\log gf$\\
\hline
Ca~{\sc ii}& 3933.664 &  0.000 & +0.105   & O~{\sc i}  & 6155.989 & 10.740 & $-$1.161 \\
Sr~{\sc ii}& 4215.519 &  0.000 & +0.145   & O~{\sc i}  & 6156.737 & 10.740 & $-$1.521 \\
Sc~{\sc ii}& 4320.732 &  0.605 & $-$0.260 & O~{\sc i}  & 6156.755 & 10.740 & $-$0.931 \\
Fe~{\sc i} & 6137.694 &  2.588 & $-$1.403 & O~{\sc i}  & 6156.778 & 10.740 & $-$0.731 \\
Ba~{\sc ii}& 6142.928 &  0.552 & $-$0.992 & O~{\sc i}  & 6158.149 & 10.741 & $-$1.891 \\
Si~{\sc i} & 6143.125 &  5.964 & $-$2.790 & O~{\sc i}  & 6158.172 & 10.741 & $-$1.031 \\
Si~{\sc i} & 6145.016 &  5.616 & $-$0.820 & O~{\sc i}  & 6158.187 & 10.741 & $-$0.441 \\
Fe~{\sc ii}& 6147.741 &  3.889 & $-$2.721 & Ca~{\sc i} & 6161.297 &  2.523 & $-$1.020 \\
Fe~{\sc ii}& 6149.258 &  3.889 & $-$2.724 & Ca~{\sc i} & 6162.173 &  1.899 &  +0.100  \\
Fe~{\sc i} & 6151.617 &  2.176 & $-$3.299 & Ca~{\sc i} & 6163.755 &  2.521 & $-$1.020 \\
Si~{\sc i} & 6155.134 &  5.619 & $-$0.400 & Fe~{\sc i} & 6165.361 &  4.143 & $-$1.550 \\
O~{\sc i}  & 6155.961 & 10.740 & $-$1.401 & Ca~{\sc i} & 6166.439 &  2.521 & $-$0.900 \\
O~{\sc i}  & 6155.971 & 10.740 & $-$1.051 &            &          &        &          \\
\hline
\end{tabular}
\end{center}
Given are the species,  air wavelength (in \AA), lower excitation potential (in eV),
and logarithm of statistical weight (of the lower level) times absorption oscillator strength
(in dex).
\end{table}

\setcounter{table}{3}
\begin{table}[h]
\caption{Resulting abundances and corresponding equivalent widths.}
\scriptsize
\begin{center}
\begin{tabular}{cccrcrcrcccccr}\hline\hline
\hline
Sp.code & Phase & $A$(Ca) & $W_{3934}$ & $A$(Sr) & $W_{4215}$ & $A$(Sc) & $W_{4320}$ & 
$A$(O)  & $A$(Si) & $A$(Ca) & $A$(Ba) & $A$(Fe) & $v_{\rm e}\sin i$\\
        &       &  (dex)  & (m\AA) &  (dex)  & (m\AA)  &  (dex)  & (m\AA) & 
(dex)   & (dex) & (dex)& (dex) & (dex) & (km~s$^{-1}$) \\ 
\hline
      &    & \multicolumn{2}{c}{[Ca~{\sc ii} 3934]} &  \multicolumn{2}{c}{[Sr~{\sc ii} 4215]} &
 \multicolumn{2}{c}{[Sc~{\sc ii} 4320]}  &  \multicolumn{6}{c}{[Solutions from the 6130--6180~\AA\ region fitting]} \\
1218B  & 0.000  & 5.82  & 1877  & 3.22  & 248  & 3.33  & 130  & 8.70  & 7.20  & 5.93  & 3.37  & 7.67  & 85.2\\
1220A  & 0.032  & 5.82  & 1879  & 3.16  & 244  & 3.35  & 132  & 8.67  & 7.19  & 5.91  & 3.37  & 7.67  & 85.2\\
1218C  & 0.180  & 5.84  & 1914  & 2.92  & 226  & 3.33  & 130  & 8.69  & 7.13  & 5.93  & 3.29  & 7.62  & 87.7\\
1220B  & 0.211  & 5.80  & 1826  & 2.99  & 231  & 3.42  & 137  & 8.67  & 7.13  & 5.90  & 3.31  & 7.63  & 88.7\\
1220C  & 0.360  & 5.90  & 2057  & 3.35  & 258  & 3.43  & 138  & 8.69  & 7.12  & 5.89  & 3.31  & 7.64  & 87.0\\
1214A  & 0.472  & 5.82  & 1877  & 3.25  & 250  & 3.32  & 129  & 8.68  & 7.20  & 5.94  & 3.37  & 7.65  & 85.1\\
1216A  & 0.649  & 5.78  & 1795  & 3.35  & 258  & 3.35  & 132  & 8.69  & 7.16  & 5.89  & 3.30  & 7.64  & 87.2\\
1214B  & 0.719  & 5.77  & 1771  & 3.21  & 248  & 3.37  & 134  & 8.70  & 7.15  & 5.89  & 3.33  & 7.64  & 89.2\\
1215A  & 0.748  & 5.76  & 1755  & 3.22  & 248  & 3.28  & 125  & 8.69  & 7.13  & 5.88  & 3.23  & 7.62  & 87.4\\
1218A  & 0.823  & 5.77  & 1775  & 3.25  & 250  & 3.28  & 125  & 8.68  & 7.08  & 5.87  & 3.31  & 7.63  & 86.3\\
\hline
       &  (mean) & 5.81  & 1853  & 3.19  & 246  & 3.35  & 131  & 8.68  & 7.15  & 5.90  & 3.32  & 7.64  & 86.9\\
       & (std.dev.)  & 0.04  &   90  & 0.14  &  10  & 0.05  &   4  & 0.01  & 0.04  & 0.02  & 0.04  & 0.02  &  1.4\\
\hline
\end{tabular}
\end{center}
Given are the abundances ($A$) derived from the spectrum fitting analysis in each of the 4 regions, 
and the corresponding equivalent widths ($W$) of Ca~{\sc ii} 3934, Sr~{\sc ii} 4215, and Sc~{\sc ii} 4320 
lines inversely derived from the abundance solutions. The mean values and their standard deviations are 
also presented in the last two rows. As usual, $A$(X) is the logarithmic number abundance of element X 
normalized with respect to H as $A$(H) = 12.00.
\end{table}

\setcounter{table}{4}
\begin{table}[h]
\caption{Variations of equivalent widths in response to changing $T_{\rm eff}$ or $\log g$.}
\scriptsize
\begin{center}
\begin{tabular}{ccrrrrrcccc}\hline\hline
\hline
Line  & $A_{\rm given}$  & $W_{\rm std}$  &  $W_{T+}$  & $W_{T-}$  & $W_{g+}$ & $W_{g-}$ & $\delta_{T+}$ 
& $\delta_{T-}$   & $\delta_{g+}$  & $\delta_{g-}$  \\
\hline
Ca~{\sc ii}~3934  & 5.807  & 1851.6  & 1527.2  & 2231.2  & 1855.5  & 1848.7  & $-0.175$  & +0.205  & +0.002  & $-0.002$ \\
Sr~{\sc ii}~4215  & 3.192  &  245.9  &  229.0  &  261.9  &  245.6  &  246.2  & $-0.069$  & +0.065  & $-0.001$  & +0.001 \\
Sc~{\sc ii}~4320  & 3.345  &  131.2  &  117.7  &  143.4  &  129.9  &  132.6  & $-0.103$  & +0.093  & $-0.010$  & +0.011 \\
\hline
\end{tabular}
\end{center}
Given in columns 3--7 are the equivalent widths in m\AA\ (for the given abundance 
$A_{\rm give}$, which is the mean of the 10 $A$ values for different phases; cf. Table~4) 
calculated for the standard model atmosphere ($T_{\rm eff} = 8371$~K, $\log g = 4.22$) 
and four model atmospheres in which $T_{\rm eff}$ or $\log g$ is interchangeably 
varied from the standard values (suffixes `$T+$', `$T-$', `$g+$', and `$g-$' denote 
the cases of $\Delta T_{\rm eff} = +250$~K,  $\Delta T_{\rm eff} = -250$~K,
$\Delta\log g = +0.05$, and $\Delta\log g =-0.05$, respectively).
The relative changes of $W$ are also presented in the following columns 8--11
($\delta _{T+} \equiv (W_{T+} - W_{\rm std})/W_{\rm std}$, etc.). 
\end{table}

\end{document}